\documentclass[aps,prd,twocolumn,showpacs,preprintnumbers,amsmath,amssymb]{revtex4}

\usepackage{amsmath,amssymb}

\usepackage{color}
\usepackage[dvips]{graphicx}
\usepackage{subfigure}

\newcommand{\lsim}{\mathrel{\mathop{\kern 0pt \rlap
  {\raise.2ex\hbox{$<$}}}
  \lower.9ex\hbox{\kern-.190em $\sim$}}}
\newcommand{\gsim}{\mathrel{\mathop{\kern 0pt \rlap
  {\raise.2ex\hbox{$>$}}}
  \lower.9ex\hbox{\kern-.190em $\sim$}}}

\newcommand{\beq}{\begin{equation}}
\newcommand{\eeq}{\end{equation}}
\newcommand{\bea}{\begin{eqnarray}}
\newcommand{\ena}{\end{eqnarray}}

\newcommand{\into}{\rightarrow}
        
\newcommand{\be}{\begin{equation}}
\newcommand{\ee}{\end{equation}}
\newcommand{\ba}{\begin{array}}
\newcommand{\ea}{\end{array}}

\newcommand{\eea}{\end{eqnarray}}
\newcommand{\sigmavz}{\langle \sigma v \rangle_0}
\begin{document}

\preprint{DFTT26/2006}

\title{Additional bounds on the pre Big--Bang--Nucleosynthesis Expansion by means of
$\gamma$-rays from the Galactic Center}



%
\author{F. Donato}
 \affiliation{Universit\`a degli Studi di Torino and \\
Istituto Nazionale di Fisica Nucleare, Sezione di Torino
\\ via P. Giuria 1, I--10125 Torino, Italy \\ {\tt (donato@to.infn.it)}}
\author{N. Fornengo}
\affiliation{Universit\`a degli Studi di Torino and \\
Istituto Nazionale di Fisica Nucleare, Sezione di Torino
\\ via P. Giuria 1, I--10125 Torino, Italy \\ {\tt (fornengo@to.infn.it)}}
\author{M. Schelke}
\affiliation{Istituto Nazionale di Fisica Nucleare, Sezione di Torino
\\ via P. Giuria 1, I--10125 Torino, Italy \\ {\tt (schelke@to.infn.it)}}

\date{\today}

\begin{abstract} The possibility to use $\gamma$--ray data from the Galactic Center
(GC) to constrain the cosmological evolution of the Universe in a phase prior to
primordial nucleosyntesis, namely around the time of cold dark matter (CDM)
decoupling, is analyzed. The basic idea is that in a modified cosmological scenario,
where the Hubble expansion rate is enhanced with respect to the standard case, the
CDM decoupling is anticipated and the relic abundance of  a given dark matter (DM)
candidate enhanced. This implies that the present amount of CDM in the Universe may
be explained by a  Weakly Interacting Massive Particle (WIMP) which possesses
annihilation cross section (much) larger than in standard cosmology. This enhanced
annihilation implies larger fluxes of indirect detection signals of CDM. We show that
the HESS measurements can set  bounds for WIMPs heavier than a few hundreds of GeV,
depending on the actual DM halo profile. These results are complementary to those
obtained in a previous analysis based on cosmic antiprotons.  For a Moore DM 
profile, $\gamma$--ray data limit the maximal Hubble rate enhancement to be below a
factor of 100. Moreover, a WIMP heavier than 1 TeV is not compatible with a
cosmological scenario with enhanced expansion rate prior to Big Bang Nucleosynthesis
(BBN). Less steep DM  profiles provide less stringent bounds, depending of the
cosmological scenario.
\end{abstract}


\maketitle

\section{Introduction}
\label{sec:intro}

In a recent paper \cite{pbpaper} we discussed the possibility to derive limits to the
expansion rate of the Universe around the time when cold dark matter (CDM) decouples
from the thermal bath, by using the fact that today these dark matter (DM) particles form
the halo of our Galaxy and they may annihilate producing antiprotons. 
The measured antiproton flux has been shown to be compatible with the expected background
originated by standard cosmic--ray spallation: this fact leads to the possibility to
use antiprotons as a powerful tool for constraining DM properties (see
{\em e.g.} Refs. \cite{pbar,pbarlight}). By assuming that DM is in the
form of a generic Weakly Interacting Massive Particle (WIMP) candidate, 
we obtained relatively stringent bound on the Universe
dynamics in a period prior to the primordial nucleosynthesis phase (preBBN), which
is  not directly constrained by other observations. We showed that, despite
the large uncertainties in the knowledge of the  galactic propagation of antiprotons
\cite{pbar}, bounds on the Hubble rate enhancement ranging  from a factor of a few
to a factor of 30 are present for DM masses lighter than 100 GeV, while for a mass of
500 GeV the bound falls  in the range 50--500. These bounds loosen for heavier 
DM particles. 

We remind here that the possibility to set a bound to the enhancement of the Hubble expansion  rate
in the early Universe by means of WIMP indirect detection signals relies on the fact that a larger
Hubble rate induces an anticipated DM decoupling and an ensuing larger relic abundance, for a
fixed annihilation cross section. This may happen, for instance, in scalar--tensor cosmology
\cite{scalartensorrelic}, in  quintessence models with a kination phase \cite{kination}, and for
anisotropic expansion and other models of modified expansion \cite{Barrow,kamionkowskiturner}. 

When the decoupling is anticipated, a DM relic abundance able to explain the current observational
data is obtained for larger DM annihilation cross sections, as compared to the standard
cosmological case. This means that larger indirect detection signals (which are proportional to the
DM  annihilation cross section) are predicted: the larger the enhancement of the Hubble rate, the
larger the indirect detection signal \cite{scalartensorrelic}. Comparison with experimental data
reflects in limits on the DM possible Hubble rate enhancement. The actual results depend also on the
specific cosmological model:  in Ref. \cite{pbpaper} we considered a general parametrization of the
Hubble rate temperature dependence, and for definiteness we studied different classes of models: a
Randall--Sundrum brane cosmology scenario \cite{Randall}, a kination scenario \cite{kination}, a
scalar--tensor cosmology scenario \cite{scalartensorrelic} and a simple case where the Hubble rate is
just boosted by a constant factor. The same modellizations will be considered  in the present
study.

In this paper we extend our analysis by including also the possibility to use
the data on $\gamma$--rays coming from the GC.  DM particles may
annihilate also to photons and the current experimental observations on the
$\gamma$--ray flux, especially the ones coming from the GC where a large
concentration of DM is expected, may be useful for our scope. With
$\gamma$--rays we have to face the large uncertainty coming from the DM
halo profile toward the GC, which is largely unknown. 
Nevertheless we will show that for typical DM halo
profiles, bounds to the preBBN Hubble rate may be set. 

The experimental data we will  use are those from the GC obtained by the HESS
\cite{hess2004,hess2006} and EGRET \cite{egret} telescopes. Since the HESS data refer to large
energies (above few hundreds of GeV),   we will be able to derive bounds for large-mass DM (above few
hundreds of GeV): this nicely complements our previous results on antiprotons \cite{pbpaper}, which
were quite stringent for low mass WIMPs, but loose for heavy ones. We will also show that EGRET
data in the GeV region are able to set limits in
the intermediate WIMP mass range (at least for the most steep DM profiles). 

Our analysis will be
independent of the specific DM candidate: the only assumption is that the WIMP we are considering is
responsible for the observed DM amount \cite{wmap}: 
\be
0.092\le\Omega_{\rm CDM } h^2 \le 0.124
\label{oh2 constraint}
\ee
where $\Omega_{\rm CDM }$ is the density parameter of CDM, and its annihilation cross
section is dominantly temperature--independent (or $s$--wave). Deviations from this situation will
lead to changes similar to those discussed in Ref. \cite{pbpaper} for the antiproton case, to which
we refer for additional discussion.

\section{Models with increased pre-BBN expansion}
\begin{figure}[t] \centering
\vspace{-20pt}
\includegraphics[width=1.0\columnwidth]{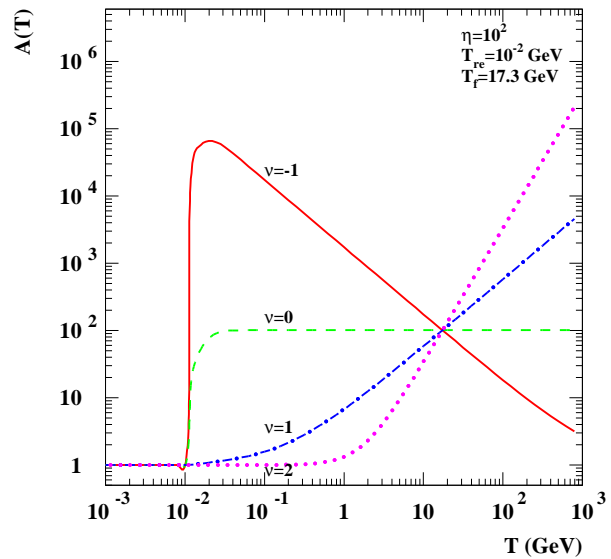}
\vspace{-20pt}
  \caption{Different models for the Hubble rate enhancement function $A(T)$, 
  defined in Eq.~(\ref{A def}). The Figure is taken from 
  Ref.~\cite{pbpaper}. Notice that the evolution of the Universe runs from right to
  left. The solid (red) line has a slope parameter $\nu = -1$, the
  dashed (green) line $\nu = 0 $, the dash-dotted (blue) $\nu = 1$ and the
  dotted (purple) line has $\nu = 2$. All curves refer to 
  $\eta = 10^2$, $T_\textrm{re} = 10^{-2}$ GeV and 
  $T_\textrm{f} = 17.3$ GeV (this freeze--out temperature refers, {\rm e.g.}, 
  to a particle with mass of 500 GeV and annihilation cross section
  $\langle\sigma_{\textrm{ann}}v\rangle =
  10^{-7}\,\textrm{GeV}^{-2}$).}
\label{fig:enhancem_func}
\end{figure} 

As mentioned in Sect. \ref{sec:intro}, it is a common feature of some  
cosmological models to predict that the expansion rate $H(T)$
in the early Universe is larger than the Hubble expansion rate $H_\textrm{GR}(T)$ 
in standard cosmology. Quite generically, 
we may introduce a function $A(T)$ to quantify the enhancement of the Hubble rate:
\bea
\label{h.Eq.a.h.gr}
H(T)&=&A(T)H_{\textrm{{GR}}}(T)\qquad\qquad\textrm{at early times}\\
H(T)&=&H_{\textrm{{GR}}}(T)\qquad\qquad\qquad\, \textrm{at later~times}. 
\eea
In Ref.~\cite{pbpaper} we introduced a parametrization of the enhancement function 
$A(T)$ which was shown to be applicable for important models like some scalar-tensor 
gravity models, some models with a kination phase and also some specific brane-world 
model. In this paper, we are going to consider the same parametrization, {\em{i.e.}}:
\be
A(T)=1+\eta\left(\frac{T}{T_\textrm{f}}\right)^\nu
  \tanh\left(
 \frac{T-T_{\textrm{re}}}{T_{\textrm{re}}}\right)
\label{A def}
\ee
for temperatures $T>T_\textrm{re}$ and $A(T) = 1$ for $T \leq T_\textrm{re}$. The 
enhancement function is shown in Fig.~\ref{fig:enhancem_func} for some 
specific parameter choices. The hyperbolic tangent serves 
to assure that $A(T)$ goes continuously to 
``1", and $H\into H_\textrm{GR}$, before some ``\emph{re-entering}" temperature, 
$T_\textrm{re}$. We must require $T_\textrm{re}\gsim 1 \, \textrm{MeV}$ to make 
sure not to be in conflict with the predictions of BBN and 
the formation of the Cosmic Microwave Background (CMB). For $T \gg T_\textrm{re}$ 
(and $\eta \gg 1$) we have approximately that:
\be
 A(T) \sim \eta \left(\frac{T}{T_\textrm{f}}\right)^\nu.
\ee
Thus, $T_\textrm{f}$ is the normalization temperature at which $A(T_\textrm{f}) 
= \eta$. As in Ref.~\cite{pbpaper} we take $T_\textrm{f}$ to be the temperature at 
which the WIMP DM candidate freezes out in standard cosmology. This means that 
$\eta$, as defined in our parametrization, is the enhancement of the Hubble rate at 
the time of the WIMP freeze--out. We will derive our results as bounds on $\eta$ for different cosmological models, characterized by the temperature--evolutionary parameter $\nu$: $\nu = 2$ refers to the Hubble rate evolution in a
Randall--Sundrum type II brane cosmology scenario of Ref. \cite{Randall};  
$\nu = 1$ is the typical kination evolution, discussed {\em e.g.} in Ref. \cite{kination});
$\nu = -1$ is representative of the behavior found in scalar--tensor cosmology in
Ref. \cite{scalartensorrelic}. The trivial case $\nu=0$ refers to an overall boost
of the Hubble rate.

For more details on the modified cosmological scenarios, the calculation of
the relic abundance in these models, including some analytical results
and discussion, we refer the reader to Ref. \cite{pbpaper}.

\section{The $\gamma$--ray signal from the GC}
\label{sec:gamma}
\begin{table*}[t]
\begin{tabular}{|c|c|c|c|c|} \hline
Isothermal &  NFW &  log--slope  & $\alpha=1.2$ & Moore  \\ 
\hline 18.9 & 6892 & 10229 & 98743 & $7.7\cdot 10^6$ \\ \hline
\end{tabular}
\caption{Values for $I_{\Delta\psi}$ in Eq.(\ref{eq:geometry}) (in units of
GeV$^2$ cm$^{-6}$ kpc). See text for details. }
\label{table:geometry}
\end{table*}

The most recent observations of the GC have been performed by the HESS
Collaboration in  the hundreds of GeV - few tens of TeV energy range.  In Refs.
\cite{hess2004,hess2004_web,hess2006} they have reported on the spectrum of very high energy
$\gamma$-rays from a point-like source in the GC,  with an unprecedented spatial
resolution, going down to a solid angle of  about $10^{-5}$ sr.  The HESS spectrum of the
central $\gamma$-ray source exhibits a clear power-law shape, with a spectral index  of
2.2\cite{hess2004}-2.25\cite{hess2006}. Diffuse $\gamma$-ray emission extended along the
galactic plane has been reported by the same Collaboration in Ref. \cite{hess_diffused}. 
The measured spectrum follows a power-law with spectral index near to 2.30 and has been
shown to be compatible with a source of locally accelerated protons interacting with 
giant molecular clouds which are extended both in longitude and in latitude. 
This diffused component  contributes to the central source only for a small
fraction (10-15\%) \cite{hess_diffused,hess2006}.  
\\
The GC hosts more than one potential  sources of $\gamma$-rays, whose nature  is still not
clear. The most motivated astrophysical sources rely on particle  acceleration near the
supermassive black hole Sgr A$^*$ located at the center of our Galaxy, or in the region of
the supernova remnant Sgr A.  The $\gamma$ radiation is produced from accelerated  charged
particles (mostly protons) interacting with the ambient matter or radiation. Another
intriguing possibility resides in the $\gamma$-ray emission  resulting from the
annihilation of DM particles.  Cosmological simulations of hierarchical
structure formation predict  a significant density cusp in the central parts of the
galaxies.  In that region annihilation of DM  particles would be strongly enhanced,  with
extraordinary high expected  fluxes for the annihilation products, such as 
$\gamma$-rays.  The angular region explored by HESS is of the same order of the one of the
probable black hole, or of the DM  cusp.

We consider here a generic WIMP which composes the galactic DM. 
The flux  $\Phi_\gamma(E_\gamma, \psi)$ of $\gamma$-rays of energy  
$E_\gamma$ originated from the WIMP pair annihilation and
coming from the angular direction $\psi$ is given by:
\begin{equation}
\Phi_\gamma(E_\gamma, \psi) = \frac{1}{4\pi} \frac{\langle\sigma
 v\rangle_0}{m_\chi^2} \frac{dN_\gamma}{d E_{\gamma}}
\frac{1}{2}I(\psi)
\label{eq:flux_gamma}
\end{equation}
where $\langle\sigma  v\rangle_0$ is the present annihilation cross section times the relative
velocity averaged over the galactic velocity distribution function.
$dN_\gamma/d E_{\gamma}$ is the energy spectrum of $\gamma$-rays
originated from a single WIMP pair annihilation and has been
calculated by means of a Monte Carlo simulations with the PYTHIA
package \cite{pythia} as described in Ref. \cite{indirect_light}.
For definiteness, as we have done also for the antiproton analysis
of Ref. \cite{pbpaper}, we are assuming the $\gamma$--ray energy spectrum originated
by a $\bar{b}b$ quark pair. A different annihilation final state will
not change substantially our results, much less than the astrophysical
uncertainties.

The quantity $I(\psi)$ is the contribution of the squared DM 
density distribution along the line of sight (l.o.s.):
\begin{equation}
I(\psi) = \int_{\rm l.o.s} \rho^2(r(\lambda,\psi))~ d\lambda(\psi).
\label{eq:los}
\end{equation}
Here $\psi$ is the angle between the l.o.s. and the line
pointing toward the GC ($\cos\psi = \cos l \cos b$, $l$ and 
$b$ being the galactic longitude and latitude, respectively). 
If Eq.(\ref{eq:los}) is used for comparison with experimental data, 
it must be averaged over the telescope observing angle $\Delta \psi$:
\beq I_{\Delta\psi} =
\frac{1}{\Delta\psi}\int_{\Delta\psi} I(\psi)~d\psi.
\label{eq:geometry}
\eeq

The geometric factor $I(\psi)$ depends quadratically 
on the DM  density profile and is very sensitive to 
its features, especially in the GC region where predictions 
for $\rho(\vec r)$ differ mostly. 
The most common spherically--averaged density profiles can be 
parametrized as:
\begin{equation}
\rho(r) = \rho_l \left(\frac{R_\odot}{r}\right)^\gamma
\left[\frac{1+(R_\odot/a)^\alpha}{1+(r/a)^\alpha}\right]^{(\beta
-\gamma)/\alpha},  
\label{eq:density}
\end{equation}
where $r=|\vec r|$, $R_\odot=8$ kpc  is the distance of the
Solar System from the GC along the galactic plane, $a$ is a scale
length and $\rho_l$ is the total local (solar neighborhood) DM density. 
In particular, the isothermal, cored density profile
is obtained with $(\alpha, \beta, \gamma)$ = (2,2,0), the Navarro, Frenk
and White (NFW) profile \cite{nfw} with $(\alpha,
\beta,\gamma)$ = (1,3,1) and the Moore et al.  profile \cite{moore} with
$(\alpha,\beta, \gamma)$ = (1.5,3,1.5). 
We also consider a non--singular DM 
density distribution function derived from extensive numerical 
simulations, whose asymptotic regime is well fitted by a 
logarithmic slope \cite{navarro}:
\begin{equation}
 \rho(r) = \rho_{-2} \; \exp\left\{ -\frac{2}{\alpha}
 \left[\left(\frac{r}{r_{-2}}\right)^\alpha -1\right] \right\},
 \label{eq:alpha}
\end{equation}
where $r_{-2}$ is the radius where the logarithmic slope is $\delta = -2$, and $\rho_{-2}
\equiv \rho(r_{-2})$.  The DM  density predicted by these profiles at the GC (for very
small $r$) varies so strongly that  also the predicted DM  signals  may  differ by
several orders of magnitude. 
The profile as steep as 1.5 is disfavored by the most
recent numerical simulations, which seem to indicate a power law index  not
exceeding 1.2 \cite{diemand04,diemand04_0}. We notice, however, that 
the experiments considered in this paper have a spatial resolution which 
is much narrower that the typical resolution size of numerical simulations. 
The simulated DM  densities in the GCs are thus mere extrapolations. 
Moreover, we must be aware of the fact that the results from many rotational curves for
galaxies of different morphological types are hardly explained by central
DM  cusps. Instead, they are more easily fitted by cored DM  distributions,
flattened towards the central region of the galaxy. 

In Table \ref{table:geometry} we show the values of the geometrical factor
$I_{\Delta\psi}$ for the HESS telescope aperture  ($\Delta\psi=10^{-5}$ sr) and for
various DM  density profiles.  The first column refers to an isothermal density
distribution with a core  $a=3.5$ kpc and the second to a NFW with a scale length $a=25$
kpc. The third is for the log-slope of Eq. (\ref{eq:alpha}) with the parameters of the 
distribution G1 in Ref. \cite{navarro}:  $\alpha=0.142$, $r_{-2}=26.4$ kpc and
$\rho_2=0.035$ GeV cm$^{-3}$.  The fourth column is the result for a profile obtained
with  $(\alpha,\beta, \gamma)$ = (1.2,3,1) with $a$=25 kpc \cite{diemand04,diemand04_0}
and  the last column refers to  a Moore et al. profile with scale  length $a=30 $ kpc. 
 The value of $\rho_l$ can be determined for each density profile requiring the
compatibility with the measurements of rotational curves and the total mass of the Galaxy
\cite{bcfs}.  For definiteness, we have fixed  $\rho_l = 0.3$ GeV cm$^{-3}$ for all the
density profiles in Table  \ref{table:geometry}. We notice that the parameter $\rho_l$
enters  as a mere scaling factor in Eq. \ref{eq:flux_gamma}: the effect of varying
$\rho_l$ is therefore easily taken into account.


\section{Constraining the Hubble rate with $\gamma$--rays}
\label{sec:constraints}

In order to derive the constraint on the enhancement of the Hubble rate, we first 
find the upper bound on $\langle \sigma v\rangle_0$ determined by the $\gamma$--ray observations, for any given halo profile.

Let us first explain our analysis for the HESS observation of high 
energy $\gamma$--rays from the GC. Here we have added the expected 
WIMP signal to a background that follows a power-law, $kE^{-\Gamma}$. 
The normalization and the index of the power-law are taken as free parameters. 
For each point in the grid--scan of $(k,\Gamma)$ we find the maximum allowed value 
of $\langle \sigma v\rangle_0$, which we statistically define as the value of the cross 
section at which the reduced $\chi^2$ equals 3. 
The reduced $\chi^2$ is defined as:
\begin{equation}
  \chi^2_{\textrm{red}} = \frac{1}{N}\sum_{n=1}^D \left(
          \frac{\Phi_\chi(n) + \Phi_B(n) -\Phi_\textrm{Obs}(n)}
			{\sigma_\textrm{Obs}(n)}\right)^2
\end{equation}
where $N$ is the number of degrees of freedom, $\Phi_\chi(n)$ is the expected WIMP  flux,
as calculated using Eq.~(\ref{eq:flux_gamma}), at the energy of the $n$--th data-point 
($\Phi_\chi(n)$ obviously vanishes when $E_n \gsim m_\chi$). The background flux is 
assumed to be $\Phi_B(n) = kE_n^{-\Gamma}$, while the observed flux is 
$\Phi_\textrm{Obs}(n)$. The 1--$\sigma$ error of the observation is denoted 
$\sigma_\textrm{Obs}(n)$. For the analysis, we use the 17
data points from the HESS 2003  observation  of the GC \cite{hess2004,hess2004_web}. They
cover the energy  range 280 GeV -- 8.83 TeV. These data can therefore be used to
constrain WIMP heavier than about 300 GeV (the annihilation process in the Galaxy
occurs almost at rest: therefore there is a kinematic cut off at the WIMP mass for the
$\gamma$--ray energy). 

\begin{figure}[t] \centering
\vspace{-20pt}
\includegraphics[width=1.0\columnwidth]{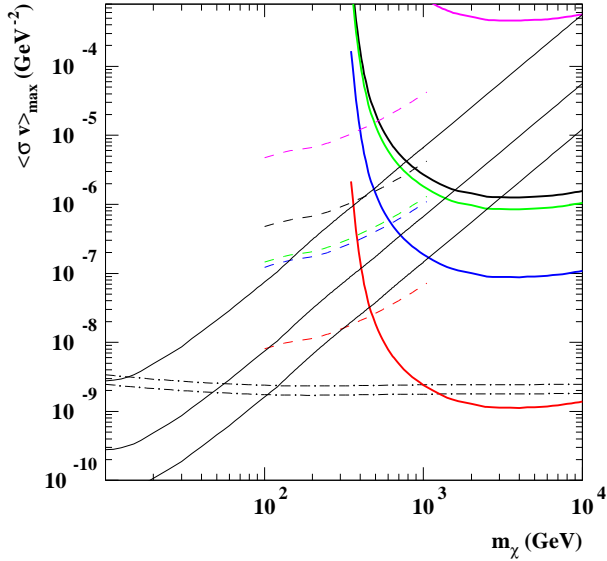}
\vspace{-20pt}
\caption{Bounds on the WIMP annihilation cross section  $\langle \sigma v\rangle_0$ as a
function of the WIMP mass. The curved lines at large WIMP mass show the {\em upper}
bounds derived from the HESS  observation of $\gamma$--rays from the GC. The derivation
has been  made for five different DM halo profiles. From top to bottom these are: the 
isothermal model, the NFW model, the 'log' slope, a power--law slope with index 1.2, the
Moore et al. profile (see text for more details).  The dashed lines show analogous {\em
upper} limits  derived using the $\gamma$--ray data from the EGRET detector, for the same
set of galactic halo models. The EGRET limits are plotted only in the mass interval which
is  relevant for the analysis of this paper. The slanted solid lines show the {\em upper}
limits coming from the observations of cosmic antiprotons \cite{pbpaper}.  The central
line refers to the best estimate for the antiproton DM signal. The upper and lower lines
refer to the astrophysical uncertainties in the galactic propagation parameters
\cite{pbpaper,pbar}. Finally, the horizontal dot--dashed lines shows the \emph{lower} 
bound on the cross section as derived from the WIMP relic density constraint assuming 
the standard cosmological model and a temperature independent WIMP annihilation cross
section.}
\label{fig:sigmalim_all}
\end{figure} 
\begin{figure}[t] \centering
\vspace{-20pt}
\includegraphics[width=1.0\columnwidth]{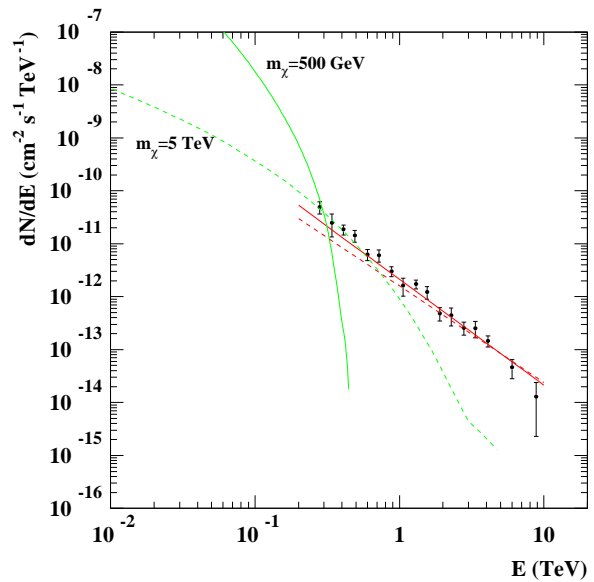}
\vspace{-20pt}
\caption{$\gamma$--ray spectra as a function of energy. The data points refer to HESS \cite{hess2004,hess2004_web}. The solid and dashed (green) curves refer to the maximal allowable
contribution to the $\gamma$--ray flux from WIMP annihilation: the cases of 500 GeV and 5 TeV
WIMPs are plotted. The solid and dashed (red) straight lines refer to a standard--source
power--law contribution to the HESS data, as obtained by our fit (see text for more details).}
\label{fig:spec}
\end{figure}
\begin{figure}[t] \centering
\vspace{-20pt}
\includegraphics[width=1.0\columnwidth]{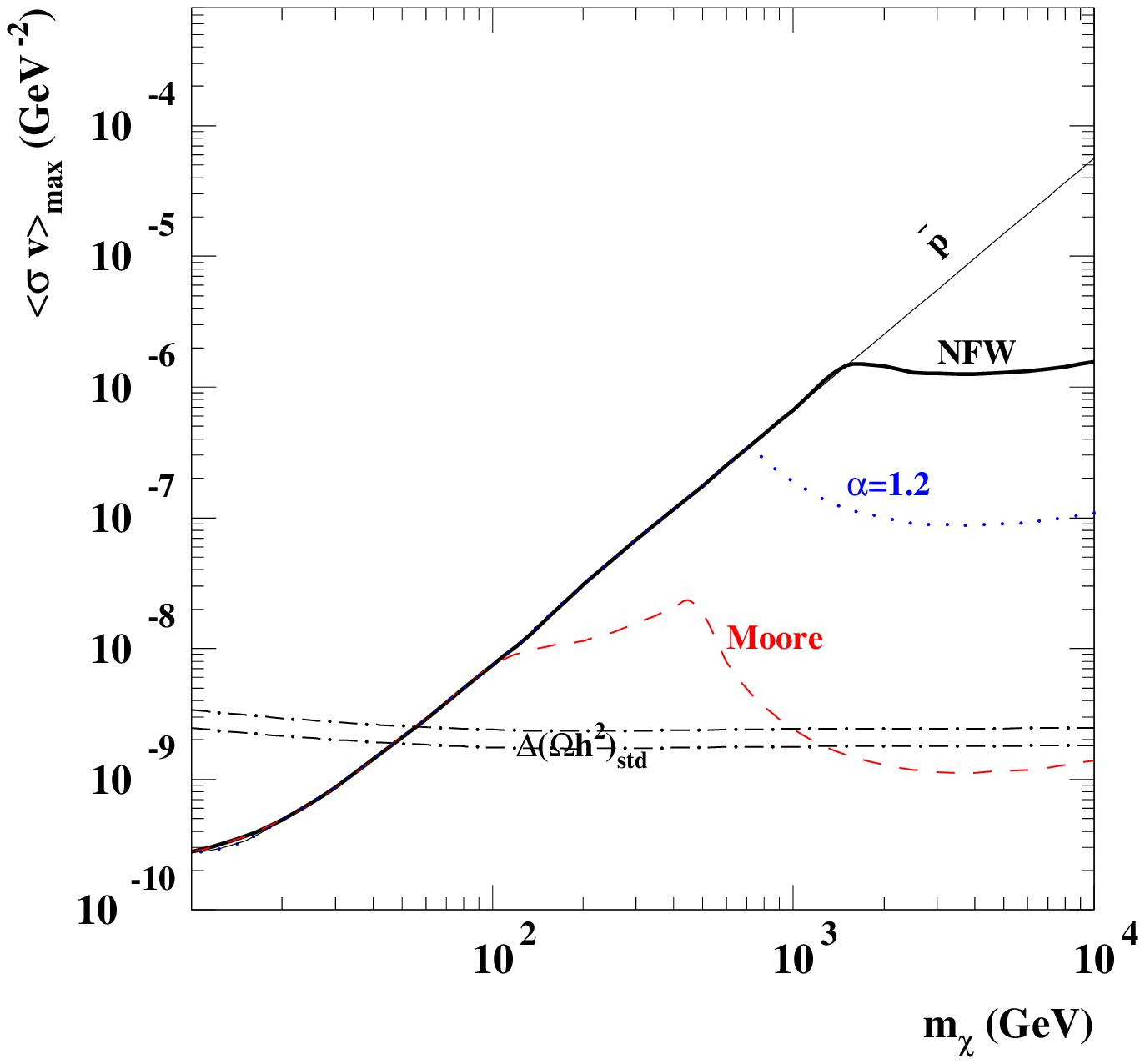}
\vspace{-20pt}
\caption{Summary of the DM indirect detection limits on the
WIMP annihilation cross section (for a temperature independent cross section).
The allowed region lies above the horizontal dot--dashed lines (which refer to
the relic density constraint) and below the slanted/curved lines, for any
given DM density profile. The slanted part of the upper bound is due to
cosmic antiprotons, while the curved part to $\gamma$--rays from the GC
(from HESS data at large WIMP masses, EGRET at intermediate masses).
}
\label{fig:sigmalim_combined}
\end{figure}

The $\chi^2$ analysis, as described above, gives us the upper limit of the cross 
section for each point in the $(k,\Gamma)$ grid for fixed WIMP mass and halo 
profile. Finally we extract the grid-point which gives the biggest value of the 
cross section. This upper bound on $\langle \sigma v\rangle_0$ is shown in 
Fig.~\ref{fig:sigmalim_all}, as a function of the WIMP 
mass for the five different halo profiles discussed above. 
As expected, the result depends strongly on the halo profile. 

Before we continue to explain how the upper limit on the cross section was 
derived from other observational data, let us show some examples of the 
differential photon production which correspond to the upper limit on the 
WIMP annihilation cross section. In Fig. \ref{fig:spec} we show the result for 
the NFW halo profile and for two different masses. The flux is calculated using 
Eq.~(\ref{eq:flux_gamma}) and inserting the upper bound on $\langle \sigma v\rangle_0$ for 
the given mass and halo profile. Also shown are the HESS 2003 observations of 
the GC as well as the fitted power-law background. Note that as the 
parameters of the background are treated as free parameters, they are different 
for the different WIMP masses and halo profiles. In the shown examples, the 
background parameters associated with the upper bound on the WIMP cross section 
are $(k,\Gamma) = (0.177 \cdot 10^{-6}~\mbox{cm$^{-2}$ s$^{-1}$ TeV$^{-1}$ sr$^{-1}$}, 2.00)$ and 
$(k,\Gamma) = (0.133 \cdot 10^{-6}~\mbox{cm$^{-2}$ s$^{-1}$ TeV$^{-1}$ sr$^{-1}$}, 1.83)$ for 
WIMP masses of 500 GeV and 5 TeV respectively.

Coming back to the limit on the WIMP annihilation cross section,
Fig. \ref{fig:sigmalim_all} shows also the upper limit
as derived from the EGRET. The EGRET data \cite{egret} span 
from energies of around 0.039 GeV to around 14.9 GeV with an 
 angular resolution given by the
longitude--latitude aperture: $|\Delta l| \leq 5^\circ$, $|\Delta b|
\leq 2^\circ$. 
The geometric factor for the EGRET experiment has been taken from Ref.
\cite{indirect_light}, for the same density distribution functions described in 
the previous Section. At these energies there is 
a $\gamma$--ray background from nucleonic reactions between cosmic rays and interstellar medium, 
electron bremsstrahlung and inverse Compton. We assume it to be the same as 
the one described in Ref. \cite{indirect_light}. 
We apply the following analysis only for the two highest EGRET energy bins, 
i.e.~$E_n \sim 6.2$ GeV and $E_n \sim 14.9$ GeV, which are the most constraining
for the masses we are dealing with. For a given WIMP mass and halo 
profile we make a scan in the WIMP annihilation cross section to find its upper 
limit taken to be the $2\sigma$ bound:
\be
 \Phi_\chi(n) + \Phi_B(n) -\Phi_\textrm{Obs}(n) \leq 
			2 \, \sigma_\textrm{Obs}(n)
\ee
The observational error, that we use, only includes the statistical error.
We use the limit from the most constraining of the two data points. For the 
mass range relevant here, it is always the highest energy bin which provides
the bound, 
except for a WIMP mass of 100 GeV. In Fig. \ref{fig:sigmalim_all} we
show only the WIMP range where the EGRET data could be of importance for the 
analysis of this paper. 
Finally, Fig. \ref{fig:sigmalim_all} shows, as slanted solid lines, the upper limit on 
$\langle \sigma v\rangle_0$ derived in our previous analysis \cite{pbpaper} of the observational data on cosmic antiprotons.
The central line refers to the best estimate for the antiproton DM
signal, {\em i.e} when the best set of astrophysical parameters are used in
the calculation of the diffusion processes for the galactic cosmic rays \cite{pbar}.  
The upper and lower lines refer to the uncertainty band arising from astrophysical uncertainties
in the galactic propagation parameters \cite{pbpaper,pbar}. The horizontal dot--dashed
line denote instead the {\em lower} limit on the annihilation cross section
derived from the cosmological bound on the WIMP relic density Eq. (\ref{oh2 constraint}) in standard cosmology.

The analysis of Fig. \ref{fig:sigmalim_all} shows interesting properties. In addition to
the already discussed bound on $\sigmavz$ from antiprotons, which sizeably constrains
the maximal allowable annihilation cross section, especially for WIMPs lighter than
a few hundreds of GeV, we now also have the bounds coming from $\gamma$--rays from the GC, which instead are relevant for heavy WIMPs. This is a consequence of the fact that HESS data refer to
an energy range from a few hundreds of GeV up to few TeV. The figure clearly shows that, in order to have
bounds on $\sigmavz$, the signal must be quite sizeable and this is possible only for very
steep DM  profiles like the Moore and NFW ones. In the case of the Moore profile, we have
a tension between the $\gamma$--rays HESS data and the cosmological limit even for standard
cosmology (a situation analogous to the one already observed for the antiproton signal
produced by light WIMPs \cite{pbpaper,pbarlight}). Should the DM  profile be the Moore
one, very little room would be allowed for $\sigmavz$. This bound would also imply a finite
possible range for the WIMP mass: from 50--100 GeV to about 1 TeV (even for standard cosmology),
as a combination of cosmological data on the amount of DM, the antiproton component
of cosmic--rays and the $\gamma$--ray signal for the GC. Less steep profiles
are clearly much less constraining, and the isothermal case is always ineffective, being always less relevant than the antiproton bound. We remind that since antiprotons diffuse in the
Galaxy, their flux is only mildly dependent on the DM  profile \cite{pbar}: the antiproton bounds
are therefore practically unaffected by the choice of different halo shapes.

A summary of the different bounds derived in our analysis is given in Fig.
\ref{fig:sigmalim_combined}, where we combine the upper limits on the WIMP  annihilation cross
section from the observations of the cosmic antiprotons and  of the GC $\gamma$--rays, as observed by
both EGRET and HESS. For any mass and halo profile we take the bound which is most constraining. The
combined upper  limit can be seen in Fig. \ref{fig:sigmalim_combined} for three halo  profiles.

\section{The maximal enhancement of the Hubble rate}

\begin{figure}[t] \centering
\vspace{-20pt}
\includegraphics[width=1.0\columnwidth]{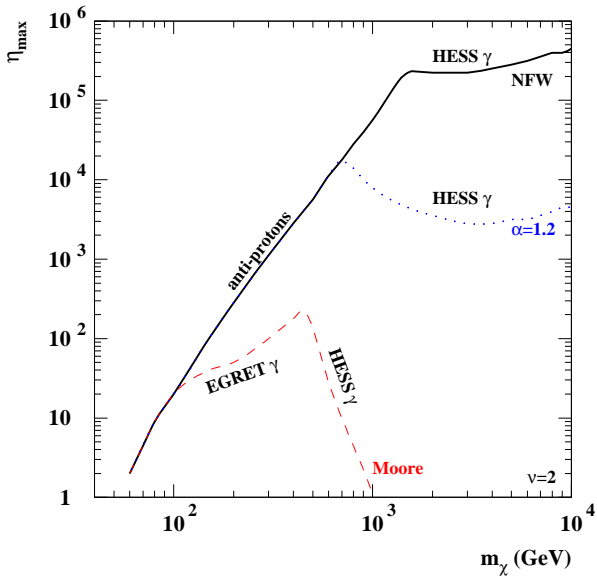}
\vspace{-20pt}\caption{Upper bound on the Hubble rate enhancement--parameter $\eta$.
The bound is 
shown as a function of the WIMP mass and has been derived by combining the 
constraints on the WIMP relic density with the constraints derived from the
observations of cosmic antiprotons and GC $\gamma$--rays. 
This figure shows the result for $T_\textrm{re} = 10^{-3}$ and $\nu = 2$ 
(RSII brane cosmology scenario \cite{Randall}), 
where $\nu$ is defined in Eq.~(\ref{A def}) for the Hubble rate enhancement 
function. We show the result for three different halo models. The labels on the
curves show the origin of the bound.}
\label{fig:etamax_nu2}
\end{figure} 

\begin{figure}[t] \centering
\vspace{-20pt}
\includegraphics[width=1.0\columnwidth]{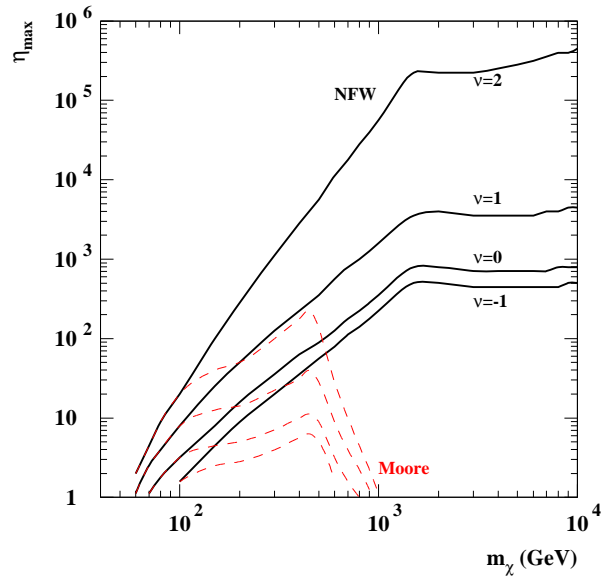}
\vspace{-20pt}
\caption{The same as in Fig. \ref{fig:etamax_nu2}, for different values of the
parameter $\nu$: $\nu = 2$ (RSII brane cosmology scenario \cite{Randall}), 
$\nu = 1$ (kination scenario \cite{kination}),
$\nu = 0$ and $\nu = -1$ (scalar--tensor cosmology scenario \cite{scalartensorrelic}).
The NFW and Moore DM density profiles are shown.}
\label{fig:etamax_allnu}
\end{figure}

An increased Hubble rate in the early Universe would increase the relic density of a given 
WIMP as compared to the situation in standard cosmology. The increase of the 
relic density is due to an anticipated freeze--out of the WIMP, as the 
annihilation rate cannot keep up with the expansion rate as long as in the 
standard case. WIMPs which satisfy the density constraint of Eq.~(\ref{oh2 constraint}) 
in the modified scenario would thus be underabundant in standard cosmology. 
Because of the inverse proportionality between the WIMP relic density and 
annihilation cross section, the WIMPs which fulfill the density constraint in 
the modified cosmologies have larger cross section than those WIMPs which 
fulfill it in the standard case. The possible enhancement of the Hubble rate can 
therefore be constrained by applying at the same time the relic density 
constraint for the DM and the upper bound on the WIMP annihilation 
cross section as derived from the indirect searches for DM.

The argument above built on the crucial assumption that the WIMP annihilation 
cross section $\langle \sigma v\rangle$ is temperature independent. If this is 
not the case then the constraint from the indirect searches, which bound  
the present $\langle \sigma v\rangle_0$, could not be directly combined with the 
constraints on the relic density, which depends on the cross section in the early 
Universe. The relation between the cross section in the two epochs should be 
taken into account, a relation which would often lead to a looser bound on the 
enhancement of the Hubble rate. A discussion of this topic was given in Ref. \cite{pbpaper},
where it was shown that modifications are usually not very large, unless some
specific situations, like {\em e.g.} coannihilation, occur. In this paper we
will show our results only for the case where the WIMP annihilation 
cross section is temperature independent.

Let us derive the bound on the enhancement of the Hubble rate from 
the combination of the constraints on the WIMP relic density and cross section. 
To obtain the WIMP relic density we solve the Boltzmann equation implemented
with the modified Hubble function, Eq.~(\ref{A def}). Let us 
therefore again go through the free parameters of the enhancement function. The 
freeze--out temperature, $T_\textrm{f}$, is determined by the WIMP mass and 
annihilation cross section and is therefore not a free parameter. The 
{\em{re--entering}} temperature, $T_\textrm{re}$, is a free parameter, but we 
showed in Ref.~\cite{pbpaper} that the bound on the Hubble expansion is 
independent of $T_\textrm{re}$ as long as $T_\textrm{re} \ll T_\textrm{f}$. In this 
paper we set $T_\textrm{re} = 10^{-3}$ GeV, which is always much lower than the 
freeze--out temperature and which is the lowest value we can assume not
to spoil BBN predictions. As we have mentioned earlier, the exponent $\nu$ in the 
enhancement function selects the kind of cosmological model. The only true free 
parameter is therefore $\eta$, which is normalized as the enhancement of the Hubble 
function at the time where the WIMP freezes out in standard cosmology. 

To derive the upper bound on the parameter $\eta$ we use for each WIMP mass 
and halo profile the upper bound on the annihilation cross section  
displayed in Fig. \ref{fig:sigmalim_combined}. For a given cosmological model, 
determined by the value of the $\nu$ parameter, the upper value of $\eta$ is 
then found where the solution of the Boltzmann equation 
satisfies the upper bound of the density constraint Eq.~(\ref{oh2 constraint}). The upper 
bound on $\eta$ as a function of the WIMP mass is shown in 
Fig. \ref{fig:etamax_nu2} for $\nu = 2$ and in Fig.~\ref{fig:etamax_allnu} also 
for $\nu = -1,0,1$. Knowing $\eta$, we can calculate the enhancement function 
$A(T)$ at any time once we choose a cosmological model.

Figs. \ref{fig:etamax_nu2} and \ref{fig:sigmalim_combined} show that $\gamma$--ray data
may be quite effective in constraining the preBBN Hubble rate for heavy WIMPs, and nicely
complement in this large--mass range the antiproton results. In order to set bounds more
stringent than antiprotons, however, a steep density profile is required: in the case of
a NFW distribution, the $\gamma$--ray observations of HESS are able to set limits only
for WIMPs heavier than 1 TeV.  For this mass range, however, depending on the actual
cosmological evolution, the bound can be relevant and much stronger than the antiproton
bound: in the case of  kination models, the $\gamma$--rays predicted for a NFW profile
limits the maximal Hubble rate enhancement to be less than a factor of 5000; for
scalar--tensor cosmologies the maximal enhancement goes down to a factor of 500. On the
other hand, for a Moore profile the bounds are quite stringent: the maximal enhancement
of the Hubble rate in this case is a factor of 100. In addition, for the Moore profile an
enhancement is not possible for all the WIMP mass  exceeding 1 TeV. 

As an example of how our limits can be further used to constrain the basic properties of
specific cosmological models, let us consider the implications for the brane
Randall--Sundrum II model \cite{Randall}, which in our notations corresponds to 
$\nu = 2$ and $\eta = \sqrt{\rho_\textrm{r}(T_\textrm{f})/(2\lambda)}$. 
Here $\rho_\textrm{r}$ is the radiation energy density and 
$\lambda$ is the tension of the  brane, related to the 5--dimensional Planck mass 
$M_5$ by  the relation $\lambda = \frac{3}{4\pi}\frac{M_5^6}{M_\textrm{pl}^2}$.
For the Moore profile and WIMP masses $m_\chi = 
\mathcal{O}(500\,\textrm{GeV})$ we see from Fig.~\ref{fig:etamax_nu2} that $\eta 
< \mathcal{O}(10^2)$. 
This implies $\lambda \gsim 2\cdot10^2\,\textrm{GeV}^4$, which corresponds to a lower
bound on $M_5$ of $M_5 \gsim 7\cdot10^3$ TeV. This is almost two orders of magnitude
better than what was found in Ref. \cite{pbpaper} by using antiproton data for the same
value of WIMP mass. It is more stringent than what can be obtained from other
cosmological tests: BBN sets the limit $M_5 > 13$ TeV \cite{Scherrer,Durrer}, while 
Ref. \cite{Nihei} sets $M_5 \gsim 600$ TeV but for a DM candidate in a specific
supersymmetric model. 
Microgravity experiments \cite{Durrer} still set the best bound $M_5 > 10^5$ TeV.


\section{Conclusions}
\label{sec:conclusion}

In this paper we have discussed the possibility to use $\gamma$--ray data
from the GC to constrain the cosmological evolution of the Universe
in a phase prior to primordial nucleosynthesis, namely around the time of
CDM decoupling from the primeval plasma. We extended the arguments
already discussed in a previous paper of ours \cite{pbpaper}, where instead cosmic--ray
antiprotons were used. The basic idea is that in a modified cosmological scenario,
where the Hubble expansion rate is enhanced with respect to the standard case, the
CDM decoupling is anticipated and therefore the relic abundance of
a given DM  candidate is enhanced. This implies that the present amount of CDM  in
the Universe may be explained by a WIMP which possesses annihilation cross section
(much) larger than in standard cosmology. This enhanced annihilation cross section
implies larger fluxes of indirect detection signals of CDM, due to the annihilation
of relic WIMPs in the halo of our Galaxy.

The stringent bounds on the maximal enhancement of the preBBN cosmic evolution, 
determined by the antiproton signal from DM  annihilation and obtained in Ref.
\cite{pbpaper}, have been complemented here by $\gamma$--ray searches. We have shown that
the HESS measurements, which refer to relatively large $\gamma$--ray energies, are able
to set constraints for WIMPs heavier than a few hundreds of GeV, depending on the actual
DM halo profile. These results are complementary to those coming from antiprotons, which
instead are important for WIMPs lighter than a few hundreds of GeV. In the case of a
Moore profile, these bounds are very strong, and imply that a WIMP heavier than about 1
TeV is not compatible with a cosmological scenario with enhanced expansion rate prior to
BBN. Less steep profiles provide less stringent bounds, always for heavy WIMPs: the NFW
halo bounds the maximal Hubble rate enhancement to be below a factor between $5\times
10^2$ and $5\times 10^5$, depending of the cosmological scenario. On the other hand, an
isothermal sphere does not provide any relevant limit (better than the antiproton bound)
for any mass. $\gamma$--ray data from the EGRET satellite are important for
intermediate--mass WIMPs, but only for the very steep Moore profile.

Data from the GLAST satellite--based experiment will add relevant information  for DM 
particles in a range of masses which goes from 100 GeV to a few hundreds of GeV,
furtherly complementing the analysis we have been able to perform by using antiprotons
(relevant for masses below 100-200 GeV) and available  $\gamma$--rays from the GC
(relevant for masses above about 500 GeV).

We can therefore conclude that DM indirect detection searches, in addition of being a powerful and
important tool for studying the DM  component of the Universe, may also have an important role in
constraining the cosmic evolution, with an impact on dark energy models, modified gravity scenarios
and theories of extra--dimensions.

\acknowledgments 
We thank Prof. A. Bottino for stimulating and fruitful discussions.
M.S. thanks P. Gondolo for useful and interesting comments. 
We acknowledge Research Grants funded jointly by the Italian Ministero
dell'Istruzione, dell'Universit\`a e della Ricerca (MIUR), by the
University of Torino and by the Istituto Nazionale di Fisica Nucleare
(INFN) within the {\sl Astroparticle Physics Project}.



\begin{thebibliography}{99}

\bibitem{pbpaper}
 Schelke M, Catena R,  Fornengo N, Masiero A, Pietroni M, (2006) 
Phys. Rev. D {\bf 74} 083505.

\bibitem{pbar}
 Donato F, Fornengo N, Maurin D, Salati P and
Taillet R, (2004)  Phys. Rev. D {\bf 69} 063501. 

\bibitem{pbarlight} 
A. Bottino, F. Donato, N. Fornengo, P. Salati,
Phys. Rev. D {\bf 72} (2005) 083518. 

\bibitem{scalartensorrelic} 
R.~Catena, N.~Fornengo, A.~Masiero,
M.~Pietroni and F.~Rosati, Phys.\ Rev.\ D {\bf 70}, 063519 (2004)


\bibitem{kination}
  P.~Salati, Phys.\ Lett.\ B {\bf 571}, 121 (2003);
  F.~Rosati, Phys.~Lett.~B {\bf 570}, 5 (2003);
  S.~Profumo and P.~Ullio, JCAP {\bf 0311}, 006 (2003);
  C.~Pallis, JCAP {\bf 0510}, 015 (2005).

\bibitem{Barrow} 
  J.~D.~Barrow, Nucl.~Phys.~B {\bf 208}, 501 (1982).

\bibitem{kamionkowskiturner}
  M.~Kamionkowski and M.~S.~Turner, Phys.~Rev.~D {\bf 42}, 3310 (1990).

\bibitem{Randall}
  L.~Randall and R.~Sundrum,
  Phys.\ Rev.\ Lett.\  {\bf 83}, 4690 (1999).
  
\bibitem{hess2004}
F. Aharonian {\it et al.} (HESS Collaboration), Astron.Astrophys. 425 (2004) L13-L17

\bibitem{hess2006}
F. Aharonian {\it et al.} (HESS Collaboration), Phys. Rev. Lett. 97 (2006) 221102

\bibitem{egret}
S.D. Hunter {\it et al.} (EGRET Collaboration), Astrophys. J. {\bf 481}, 205 (1997).

\bibitem{wmap}
Spergel, D N {\it et al.} (WMAP Collaboration), [arXiv:astro-ph/0603449], subm. Astrophys. J. 

\bibitem{hess2004_web} 
Data points have been taken from the table on the Collaboration web site:
http://www.mpi-hd.mpg.de/hfm/HESS/HESS.html

\bibitem{hess_diffused}
F. Aharonian {\it et al.} (HESS Collaboration), Nature 439 (2006) 695. 

\bibitem{pythia}
T.~Sj\"ostrand, P.~Eden, C.~Friberg, L.~Lonnblad,
G.~Miu, S.~Mrenna and E.~Norrbin,
Comput.\ Phys.\ Commun.\  {\bf 135}, 238 (2001).

\bibitem{indirect_light}
A. Bottino, F. Donato, N. Fornengo, S. Scopel, 
Phys. Rev. {\bf D70} (2004) 015005.

\bibitem{nfw} 
J.F. Navarro, C.S. Frenk and S.D.M. White, Astrophys. J. {\bf 462}, 563 (1996).

\bibitem{moore} 
Moore B {\it et al.}, Mon. Not. Roy.  Astron. Soc. {\bf 310}, 1147 (1999). 

\bibitem{navarro} 
Navarro J F {\it et al.},  (2004)  MNRAS {\bf 349} 1039. 

\bibitem{diemand04}
Diemand J, Moore B, Stadel J, (2004) MNRAS {\bf 352}  535.

\bibitem{diemand04_0}
Diemand J, Moore B, Stadel J, (2004) MNRAS {\bf 353}  624.

\bibitem{bcfs} 
P. Belli, R. Cerulli, N. Fornengo and S. Scopel,
Phys. Rev. D {\bf 66}, 043503 (2002).

\bibitem{Scherrer}
Bratt, J D, Gault, A C, Scherrer, R J, and Walker, T P, 
(2002) Phys. Lett. {\bf 546} 19. 


\bibitem{Durrer}
  R.~Durrer,   
  AIP Conf.\ Proc.\  {\bf 782}, 202 (2005)
  [arXiv:hep-th/0507006].
  
\bibitem{Nihei}
Nihei, T, Okada, T, and Seto, O, (2005) Phys. Rev. {\bf D71} 063535
  
\end{thebibliography}
\end{document}